\documentclass[twocolumn,english]{article}
\usepackage[T1]{fontenc}
\usepackage{amssymb}
\usepackage{graphicx}

\makeatletter
\usepackage{a4}
\input{epsfig.sty}
\def\fnum@table{\tablename~{\bf\thetable}}
\def\fnum@figure{\figurename~{\bf\thefigure}}
\def\tablename{\footnotesize{\bf Table}}
\def\figurename{\footnotesize{\bf Figure}}

\def\be{\begin{equation}}
\def\ee{\end{equation}}

\makeatother

\usepackage{babel}
\begin{document}

\title{\textbf{On the prompt contribution to the atmospheric neutrino flux}}

\author{Sergey Ostapchenko, Maria Vittoria Garzelli
  and G\"unter Sigl\\
\textit{\small Universit\"at Hamburg, II Institut f\"ur Theoretische
Physik, 22761 Hamburg, Germany}\\
}

\maketitle
\begin{center}
\textbf{Abstract}
\par\end{center}
The prompt contribution to the atmospheric neutrino flux is analyzed.
It is demonstrated that the corresponding
theoretical uncertainties related to perturbative
treatment of charm production, notably, the ones stemming from the low and
high $x$ behavior of parton distribution functions, can be conveniently
studied at the level of charm quark production. Additionally, we discuss
the non-perturbative contribution to the prompt neutrino flux, related to
the intrinsic charm content of the proton, and analyze its main features.

\section{Introduction\label{intro.sec}}

The detection of astrophysical neutrino fluxes by the IceCube experiment
\cite{aar13a,aar13b} paves
the way for establishing neutrino astronomy as a viable method for studying the
remote universe. Among the relevant research activities are ones aiming at
a reliable estimation of the background for such measurements, produced by
cosmic ray (CR) interactions in the atmosphere of the Earth
\cite{gaisser-book,lip93,bar04,hon07}.
Particular attention
is paid to calculations of the so-called prompt  neutrino flux
resulting from decays of charmed hadrons produced in such interactions, 
which dominates the atmospheric neutrino background for neutrino energies
$E_{\nu}\gtrsim1$  PeV \cite{tig96,pas99,enb08}.

A number of analyses have been devoted to studies of prompt neutrino production,
comparing different approaches to the problem, investigating the impact of
present uncertainties regarding parton distribution functions (PDFs) of
protons and nuclei, and studying the dependence of the results on 
the employed parametrizations of the primary CR fluxes
 \cite{bha15,gar15,gau15,bha16,ben17}.

In this work, we choose to address the problem at the level of the production
cross sections for charm (anti)quarks, using the collinear factorization
framework of the perturbative quantum chromodynamics (pQCD). We demonstrate
that the relevant fragmentation functions for charm (anti)quarks, as well as
the decay distributions for charmed hadrons, can be 
 factorized
out, such that the relevant input from pQCD is described by CR spectrum-weighted
 moments (``$Z$-factors'') of production spectra for   charm (anti)quarks.
This proves to be convenient for studying the relevant uncertainties,
notably, regarding the PDFs involved, and for specifying the kinematic regions
relevant for such calculations. 

Additionally, we discuss the non-perturbative contribution to the prompt neutrino
flux, related to the intrinsic charm content of the proton, and demonstrate
that the corresponding $Z$-factors take a particularly simple form.
However, our approach may be inapplicable to the case of
  non-perturbative charm production 
  because of potentially different hadronization mechanism in such a case.

The outline of the paper is as follows. In Section~\ref{form.sec}, we present
our formalism and derive a relation between the perturbative contribution to the
prompt atmospheric neutrino flux and the respective  $Z$-factor for charm
production. In Section~\ref{results.sec}, we present the corresponding numerical
results and analyze their dependence on the gluon PDFs in use. 
Section \ref{IC.sec} is devoted to a discussion of the intrinsic charm
contribution. Finally, we conclude in Section \ref{summary.sec}.

\section{Formalism\label{form.sec}}

The main contribution to  prompt atmospheric neutrinos is generated by  
the proton component of the primary CR flux. Concentrating, for definiteness,
on the muonic (anti)neutrinos, the relevant range of neutrino energies extends
from few hundred TeV till $\sim 10$ PeV, since for higher neutrino energies 
interactions of their would-be parent
charmed hadrons start to prevail over their decays. In turn, this involves
interactions of primary protons at energies above the so-called ``knee'' of the
CR spectrum at $E_{\rm knee}\simeq 3-4$ PeV \cite{kul59}  and below the proton ``ankle''
at $E\simeq 100$ PeV \cite{ape13a}. In that energy range, the CR proton spectrum
can be approximated by a power law   behavior,
\begin{equation}
 I_p(E_0)\simeq I_p(E_{\rm knee})\, (E_0/E_{\rm knee})^{-\gamma_p}\,, \label{eq:p-flux}
\end{equation}
  with $\gamma_p \simeq 3.1-3.3$ 
 \cite{ape13a,ant05,ape13b,ape11,aar19}.

For the corresponding prompt neutrino flux, 
one obtains \cite{gaisser-book,lip93,tig96}
\begin{eqnarray}
I_{\nu_{\mu}{\rm (prompt)}}^{(p)}(E_{\nu})
\simeq \int \!dE_0\,\frac{I_p(E_0)}{1-Z_{p{\rm -air}}^p(E_0)}
&&\nonumber \\
\times\;
\frac{dn_{p{\rm -air}}^{\nu_{\mu}{\rm (prompt)}}(E_0,E_{\nu})}
{dE_{\nu}}\,,&&
\label{eq:nu-flux}
\end{eqnarray}
where  $dn_{p{\rm -air}}^{\nu_{\mu}{\rm (prompt)}}/dE_{\nu}$ is the inclusive
spectrum of muon (anti)neutrinos resulting from decays of charmed hadrons
produced in $p$-air interactions, and $Z_{p{\rm -air}}^p$ is the
spectrum-weighted moment for proton ``regeneration'':
\begin{equation}
Z_{p{\rm -air}}^p(E)=\int \!dE_0\,\frac{I_p(E_0)}{I_p(E)}\,
\frac{dn_{p{\rm -air}}^p(E_0,E)}{dE}\,,
\label{eq:p-regener}
\end{equation}
with $dn_{p{\rm -air}}^p/dE$ being the energy distribution of secondary protons
in proton-air collisions.

For the power law primary flux (\ref{eq:p-flux}),
 Eq.\ (\ref{eq:nu-flux}) can be transformed to
\begin{eqnarray}
I_{\nu_{\mu}{\rm (prompt)}}^{(p)}(E_{\nu})\simeq I_p(E_{\rm knee})
&&\nonumber \\
\times\;\frac{(E_{\nu}/E_{\rm knee})^{-\gamma_p}}  
{1-Z_{p{\rm -air}}^p(E_{\nu},\gamma_p)}\;
 Z_{p{\rm -air}}^{\nu_{\mu}{\rm (prompt)}}(E_{\nu},\gamma_p)\,,&&
\label{eq:flux-z}
\end{eqnarray}
where we used the weak energy-dependence of the factor 
$(1-Z_{p{\rm -air}}^p)^{-1}$  \cite{tig96} to take it out of the integral, 
while the spectrum-weighted moments ($Z$-factors) $Z_{p{\rm -air}}^X$,
 $X=p,\nu_{\mu}$(prompt), are now defined as
\begin{equation}
Z_{p{\rm -air}}^X(E,\gamma_p)=\int \!dx\,x^{\gamma_p-1}\,
\frac{dn_{p{\rm -air}}^X(E/x,x)}{dx}\,.
\label{eq:X-gener}
\end{equation}
Here $dn_{p{\rm -air}}^X/dx$ is the distribution of the produced particles $X$,
with respect to the energy fraction $x=E_X/E_0$ taken  from the parent
proton. For the prompt neutrino production, it is expressed via convolutions
of the respective distributions of charmed hadrons,
 $dn_{p{\rm -air}}^{h_c}/dx_h$, with the
corresponding decay distributions, $f^{\rm dec}_{h_c\rightarrow \nu_{\mu}}$,
summed over the hadron species:
\begin{eqnarray}
\frac{dn_{p{\rm -air}}^{\nu_{\mu}{\rm (prompt)}}(E,x_{\nu})}{dx_{\nu}}=
 \sum_{h_c} \int_{x_{\nu}}^1 \! \frac{dx_h}{x_h}
&&\nonumber \\
 \times\;
\frac{dn_{p{\rm -air}}^{h_c}(E,x_h)}{dx_h}\:
 f^{\rm dec}_{h_c\rightarrow \nu_{\mu}}(x_{\nu}/x_h)\,.&&
\label{eq:sum-hadron}
\end{eqnarray}
In the high energy limit we are interested in, one can neglect the dependence
of  $f^{\rm dec}_{h\rightarrow \nu_{\mu}}$ on the hadron energy, 
while the neutrino
energy fraction $x_{\nu}/x_h$ 
is indistinguishable from
 the respective light-cone plus (LC$^+$) momentum fraction
 $(E_{\nu}+p_{z_{\nu}})/(E_h+p_{z_h})$ \cite{ko14}.

In the collinear factorization framework, $dn_{p{\rm -air}}^{h_c}/dx_h$
can be expressed via the inclusive cross section for  charm (anti)quark
production, $d\sigma_{p{\rm -air}}^{c\,(\bar c)}/dx_c$ as follows
\begin{eqnarray}
\frac{dn_{p{\rm -air}}^{h_c}(E,x_h)}{dx_h} 
=\frac{1}{\sigma_{p{\rm -air}}^{\rm inel}(E)} \sum_{c,\bar c}\int_{x_h}^1
\!\frac{dx_c}{x_c}
&&\nonumber \\
\times\; \frac{d\sigma_{p{\rm -air}}^{c\,(\bar c)}(E,x_c)}{dx_c}\,
D_{c\,(\bar c)\rightarrow h_c}(x_h/x_c)\,.&&
\label{eq:prompt}
\end{eqnarray}
Here we neglected the dependence of the charm (anti)quark fragmentation 
functions $D_{c\,(\bar c)\rightarrow h_c}$ on the factorization scale 
 for  hard parton scattering; $\sigma_{p{\rm -air}}^{\rm inel}$ is the
 inelastic proton-air cross section.

Making use of Eq.\ (\ref{eq:prompt}) in Eq.\  (\ref{eq:sum-hadron}), 
inserting the 
result into Eq.\ (\ref{eq:X-gener}), and changing to integration variables $z_{\nu}=x_{\nu}/x_c$, 
 $z_{h}=x_{h}/x_c$, we obtain
\begin{eqnarray}
Z_{p{\rm -air}}^{\nu_{\mu} {\rm (prompt)}}(E_{\nu},\gamma_p)=
 \;\int_0^1 \!dz_{\nu}\; H(z_{\nu},\gamma_p)
&&\nonumber \\
\times\;  Z_{p{\rm -air}}^{c}(E_{\nu}/z_{\nu},\gamma_p)\,.&&
\label{eq:final}
\end{eqnarray}
Here $Z_{p{\rm -air}}^{c}$ is defined by Eq.\ (\ref{eq:X-gener}), for $X=c$, and
\begin{eqnarray}
 H(z_{\nu},\gamma_p)=z_{\nu}^{\gamma_p-1}
 \sum_{c,\bar c} \sum_{h_c} \int_{z_{\nu}}^1 \!\frac{dz_{h}}{z_{h}}
&&\nonumber \\
\times\; 
D_{c\,(\bar c)\rightarrow h_c}(z_{h})\;
 f^{\rm dec}_{h_c\rightarrow \nu_{\mu}}(z_{\nu}/z_{h})\,. &&
\label{eq:zfac-c}
\end{eqnarray}

Finally, noting that small values of $z_{\nu}$ in the integrand in the
right-hand side (rhs) of Eq.\ (\ref{eq:final}) are suppressed by the factor
$z_{\nu}^{\gamma_p-1}$ [c.f., Eq.\ (\ref{eq:zfac-c})] and 
assuming that $Z_{p{\rm -air}}^{c}(E_{\nu}/z_{\nu},\gamma_p)$ changes
weakly in the relevant range of $z_{\nu}$, we get
\begin{eqnarray}
Z_{p{\rm -air}}^{\nu_{\mu} {\rm (prompt)}}(E_{\nu},\gamma_p)\simeq
 Z_{p{\rm -air}}^{c}(E_{\nu},\gamma_p)
&&\nonumber \\
\times \left[ \sum_{c,\bar c} \sum_{h_c}
 Z_{c\,(\bar c)\rightarrow h_c}^{\rm fragm}(\gamma_p)\;
 Z^{\rm dec}_{h_c\rightarrow \nu_{\mu}}(\gamma_p)\right],&&
\label{eq:final-fact}
\end{eqnarray}
with
\begin{eqnarray}
 Z_{c\,(\bar c)\rightarrow h_c}^{\rm fragm}(\gamma_p)=
 \int_0^1 \!dz\,z^{\gamma_p-1}\,
D_{c\,(\bar c)\rightarrow h_c}(z)\,, &&  \\
 Z^{\rm dec}_{h_c\rightarrow \nu_{\mu}}(\gamma_p)=
 \int_0^1 \!dz\,z^{\gamma_p-1}\,
f^{\rm dec}_{h_c\rightarrow \nu_{\mu}}(z)\,.
\label{eq:z-factors}
\end{eqnarray}

We can see that all the pQCD input in Eqs.\ (\ref{eq:final}) and  
(\ref{eq:final-fact}) is contained in the  CR spectrum-weighted
moments  $Z_{p{\rm -air}}^{c}$ of the  energy distributions of charm quarks
produced in proton-air interactions, which allows one to study the respective
uncertainties at the level of $c$-quark production.

Let us now briefly comment on the contributions of
primary nuclear species to the prompt atmospheric neutrino flux.
While  partial spectra for various nuclear mass groups of the primary CRs are
not well-determined at the energies of our interest, there are strong
experimental indications that those contain spectral breaks (``knees'')
at energies $Z_i$ times higher than the one of the proton knee,
 $Z_i$ being the characteristic charge for the $i$-th group, and the respective
 spectral slopes  $\gamma_i$ above the breaks are not too different from the proton slope
 $\gamma_p$  \cite{ape13a,ant05,ape13b,ape11,aar19,kam12,pao21}.
 To some extent, this is indeed expected,
  if all the primary species come from the same kind of 
 astrophysical sources. Adopting such a picture, partial fluxes of various
 nuclear mass groups of the primary CRs can also be described by the
 corresponding power laws,  
\begin{equation}
 I_{A_i}(E_0)\simeq I_{A_i}(Z_i\, E_{\rm knee})\:
  \left(\frac{E_0}{Z_i\,E_{\rm knee}}\right)^{-\gamma_i}\,,
 \label{eq:A-flux}
\end{equation}
$E_0$ being here the energy per nucleus.
 Further, since the prompt neutrino yield is
 intimately related to forward (high $x_c$) charm (anti)quark production,
 the so-called superposition model (see, e.g.\ Ref.\ \cite{eng92}) is fully
 applicable here: the neutrino yield from a primary nucleus of mass number
 $A_i$ and energy $E_0$ can be approximated by  $A_i$ times the yield from a
 primary proton of energy  $E_0/A_i$. This leads us to 
 [c.f., Eq.\ (\ref{eq:flux-z})]
\begin{eqnarray}
I_{\nu_{\mu}{\rm (prompt)}}^{(A_i)}(E_{\nu})\simeq 
\frac{A_i^{2-\gamma_i}\;I_{A_i}(Z_i\, E_{\rm knee})}
{1-Z_{p{\rm -air}}^p(E_{\nu},\gamma_i)}
&&\nonumber \\
\times\; \left(\frac{E_{\nu}}{Z_i\,E_{\rm knee}}\right)^{-\gamma_i}
 Z_{p{\rm -air}}^{\nu_{\mu}{\rm (prompt)}}(E_{\nu},\gamma_i)\,.&&
\label{eq:flux-A}
\end{eqnarray}
Thus, also in that case, the problem is reduced to a calculation of the CR
spectrum-weighted moments $Z_{p{\rm -air}}^{\nu_{\mu}{\rm (prompt)}}$ - this
time, using the corresponding slope $\gamma_i$ for a primary nuclear mass group
of interest. Secondly, let us recall that the relative abundances of the main
primary mass groups are of the same order of magnitude at the proton
 knee energy  (see, e.g.\ Refs.\ \cite{ant05,aar19}).
Therefore, if the primary spectral slopes for these groups are indeed similar
to the one for primary protons, $\gamma_i\simeq \gamma_p$, 
Eq.\ (\ref{eq:flux-A}) tells us that significant
contributions to the prompt atmospheric neutrino flux come from CR protons and
helium nuclei only, with the summary contribution of heavier primaries being
a $\sim 10$\% correction.

\section{Numerical results\label{results.sec}}
As the dominant contribution to charm (anti)quark production comes from the
gluon-gluon fusion process (see e.g.\ \cite{grv94})
 and the gluon PDF of a nucleus can be
approximated by a superposition of the ones of its nucleons,\footnote{Regarding
the prompt neutrino fluxes, nuclear
corrections to this approximation have been studied in Ref.\ \cite{bha16}.}
we have 
\begin{eqnarray}
 Z_{p{\rm -air}}^{c}(E,\gamma)\simeq  \int \!dx_c\; x_c^{\gamma -1}
&&\nonumber \\
\times \;  
\frac{\langle A_{\rm air}\rangle}{\sigma_{p{\rm -air}}^{\rm inel}(E/x_c)}\,
\frac{d\sigma_{pp}^{c\,(gg)}(E/x_c,x_c)}{dx_c}\,, 
\label{eq:c-gg}
\end{eqnarray}
where $\langle A_{\rm air}\rangle$ is the average mass number for air 
nuclei\footnote{In the following, we
 approximate the air composition by its most abundant element, nitrogen: 
$\langle A_{\rm air}\rangle/\sigma_{p{\rm -air}}^{\rm inel}
\simeq 14/\sigma_{pN}^{\rm inel}$, and use the predictions of the QGSJET-II
model \cite{ost11} for $\sigma_{pN}^{\rm inel}$.}
and $d\sigma_{pp}^{c\,(gg)}/dx_c$ is defined by the usual collinear
factorization ansatz:
\begin{eqnarray}
\frac{d\sigma_{pp}^{c\,(gg)}(E_p,x_c)}{dx_c}=\int \!dx^+dx^-\!
\int \!d^2k_{\perp _c}dy^{*}_c
&&\nonumber \\
\times \;\frac{d^3\hat 
\sigma_{gg\rightarrow c}(\hat s,y^{*}_c,k_{\perp _c},\mu_{\rm F}
,\mu_{\rm R})}{dy^{*}_c \,dk_{\perp _c}^2}\, g_p(x^+,\mu_{\rm F})
&&\nonumber \\
\times \; g_p(x^-,\mu_{\rm F})\:
\delta[x_c-x^+\,m_{\perp _c}\,e^{y^{*}_c}/\sqrt{\hat s}]\,.
\label{eq:sig-c}
\end{eqnarray}
Here $d^3\hat \sigma_{gg\rightarrow c}/dy^{*}_c/dk_{\perp _c}^2$ is the 
differential short distance cross section for $c$-quark production in
 the $gg$-fusion process,
for which we use the next-to-leading order (NLO) result from Ref.\ \cite{nas89},
$g_p(x,Q)$ is the gluon PDF of the proton,  
 $x^{\pm}$ are the LC$^{\pm}$ momentum fractions
for the, respectively, projectile and target gluons, 
$k_{\perp _c}$ and $y^{*}_c$ are, correspondingly, the transverse momentum and
the rapidity of the produced $c$-quark in the gluon-gluon center-of-mass (c.m.)
frame, $\hat s=x^+x^-s$ is the c.m.\ energy squared for the 
$gg$-scattering, while $s\simeq 2E_pm_p$ is the one for the proton-proton collision, $m_p$ being the proton mass.
 In the following, unless specified otherwise, we set the factorization
  $\mu_{\rm F}$ and  renormalization $\mu_{\rm R}$ scales  equal to  
   $c$-quark transverse mass 
$m_{\perp _c}=\sqrt{m_c^2+k_{\perp _c}^2}$, while using $m_c=1.3$ GeV
for the charm quark mass. In the argument of the
 $\delta$-function in  Eq.\ (\ref{eq:sig-c}), 
 we neglected the difference between the energy fraction $x_c$ of 
the $c$-quark and its LC$^+$ momentum fraction.

\begin{figure}[htb]
\includegraphics[width=0.49\textwidth]{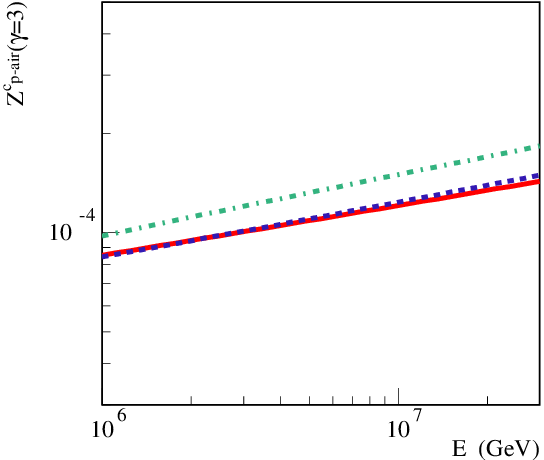}
\caption{Energy dependence of the CR spectrum-weighted moment of $c$-quark
production spectrum, $Z_{p{\rm -air}}^{c}(E,\gamma)$, for proton-air
interactions, calculated using $\gamma =3$ and employing gluon PDFs from
CT14nlo\_NF3  (solid), ABMP16\_3\_nlo (dashed),
and  NNPDF31\_nlo\_pch\_as\_0118\_nf\_3 (dashed-dotted)  PDF sets.
\label{fig:z_c}}
\end{figure}

\begin{figure*}[t]
\includegraphics[width=\textwidth]{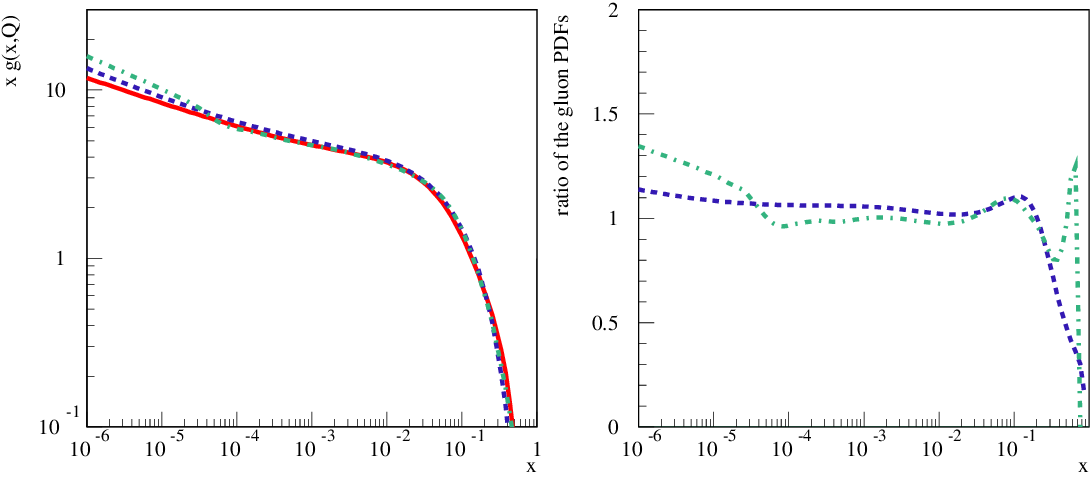}
\caption{Left: $x$-dependence of the gluon PDF $g_p(x,Q)$, for $Q=2$ GeV,
for the considered PDF sets;
the meaning of the lines is the same   as in Fig.\ \ref{fig:z_c}.
Right: the ratios of the gluon PDFs, for $Q=2$ GeV, from the  
ABMP16\_3\_nlo  and NNPDF31\_nlo\_pch\_as\_0118\_nf\_3  sets to the 
one of the CT14nlo\_NF3  PDF set -   dashed and dashed-dotted
lines, respectively.
\label{fig:g-pdf}}
\end{figure*}

In Fig.\ \ref{fig:z_c}, we plot the CR spectrum-weighted moment for charm
 production, $Z_{p{\rm -air}}^{c}(E,\gamma)$,
 calculated for $\gamma =3$, using gluon PDFs from 3-flavour NLO PDF sets
CT14nlo\_NF3 \cite{ct14}, ABMP16\_3\_nlo \cite{ale12},
and NNPDF31\_nlo\_pch\_as\_0118\_nf\_3 \cite{bal11},
as implemented in  the LHAPDF package \cite{lhapdf}.
For all the gluon PDFs employed,  we observe a similar energy dependence 
of $Z_{p{\rm -air}}^{c}$. A slightly stronger energy
rise of the $Z$-factor based on the NNPDF3.1 parametrization is due to
a somewhat steeper low-$x$ rise of the respective gluon PDF
 (see Fig.\ \ref{fig:g-pdf}).

\begin{figure*}[t]
\includegraphics[width=0.49\textwidth]{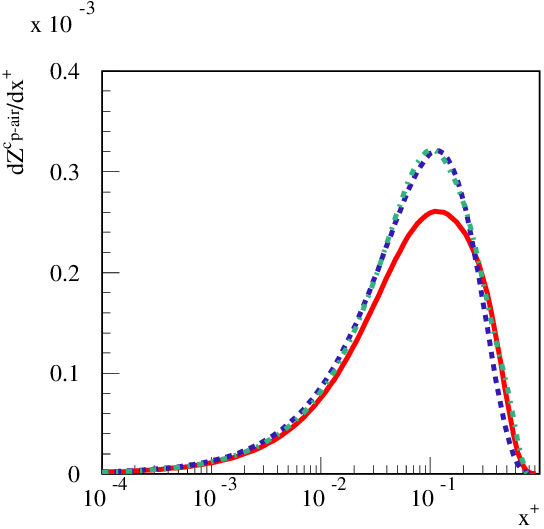}\hfill
\includegraphics[width=0.49\textwidth]{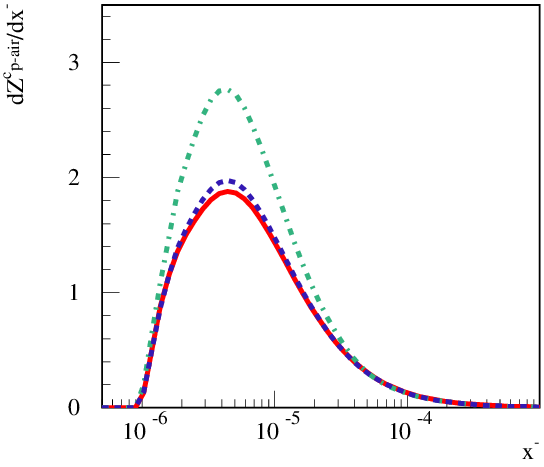}
\caption{Distributions of the  LC momentum fractions $x^{\pm}$ for,
respectively, projectile (left) and target (right) gluons,
$dZ_{p{\rm -air}}^{c}/dx^{\pm}$,  for $E=1$ PeV and $\gamma =3$.
The meaning of the lines is the same as in Fig.\ \ref{fig:z_c}.
\label{fig:dz_c/dx}}
\end{figure*}

In  Fig.\ \ref{fig:dz_c/dx}, we illustrate the range of the
 LC momentum fractions $x^{\pm}$ of the projectile and target gluons,
corresponding to maximal contributions to   $Z_{p{\rm -air}}^{c}$, for the
considered PDF sets. To this end,
we plot, for $E=1$ PeV and $\gamma =3$, 
the corresponding distributions $dZ_{p{\rm -air}}^{c}/dx^{\pm}$, as
defined by Eq.\ (\ref{eq:c-gg}), with $d\sigma_{pp}^{c\,(gg)}/dx_c$ being
replaced by the respective integrands from the rhs of Eq.\ (\ref{eq:sig-c}).
Clearly, the main contribution to  $Z_{p{\rm -air}}^{c}$ comes from
relatively high values of $x^+\sim x_c$:  since low $x_c$ is suppressed by the
factor $x_c^{\gamma -1}$ [c.f., Eqs.\  (\ref{eq:c-gg}-\ref{eq:sig-c})]. 
On the other hand, the 
target gluon PDF is mostly probed at very small values of 
$x^-\sim  m^2_c/(x^+s)\sim m_c/E$, $E$ being the charm quark energy,
as already stressed in previous studies (e.g.\ \cite{bha16}).
 Therefore, the energy rise
of $Z_{p{\rm -air}}^{c}(E,\gamma)$ is intimately related to the low-$x$ rise
of the gluon PDF $g_p(x^-,\mu_{\rm F})$, as discussed above.

\begin{figure}[htb]
\includegraphics[width=0.49\textwidth]{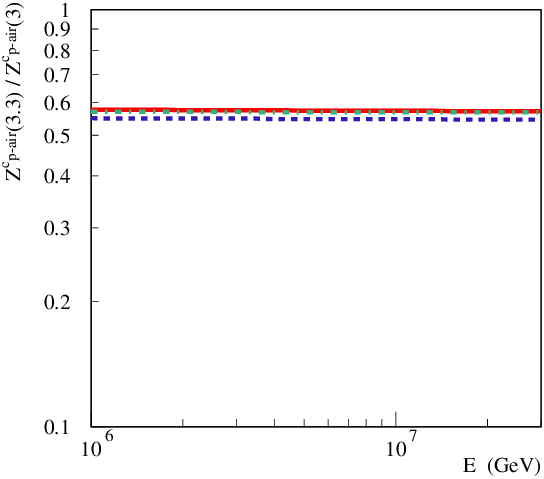}
\caption{Energy dependence of the ratio of the CR spectrum-weighted moments 
 $Z_{p{\rm -air}}^{c}(E,\gamma)$, for $\gamma =3.3$ and $\gamma =3$.
The meaning of the lines is the same   as in Fig.\ \ref{fig:z_c}.
\label{fig:z_c-slope}}
\end{figure}

\begin{figure*}[t]
\includegraphics[width=0.49\textwidth]{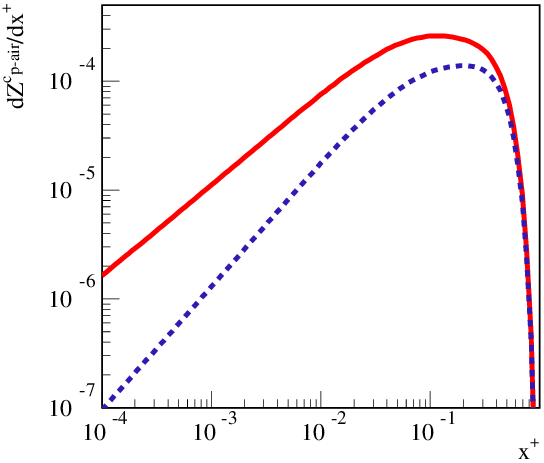}\hfill
\includegraphics[width=0.49\textwidth]{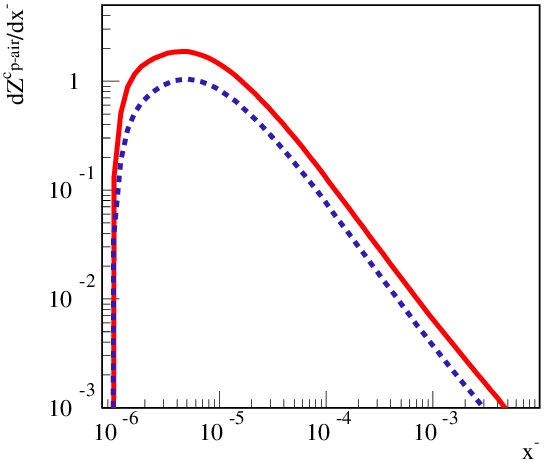}
\caption{Distributions of the  LC momentum fractions $x^{\pm}$ for,
respectively, projectile (left) and target (right) gluons,
$dZ_{p{\rm -air}}^{c}/dx^{\pm}$,  for $E=1$ PeV, using gluon PDF from the
 CT14nlo\_NF3 set. Solid lines correspond to
 $\gamma =3$ and dashed ones - to $\gamma =3.3$.
\label{fig:z_c-xpm-gamma}}
\end{figure*}

To estimate the impact of uncertainties related to the primary proton spectral
slope, we plot in  Fig.\ \ref{fig:z_c-slope} the energy dependence of the
 ratio $Z_{p{\rm -air}}^{c}(E,\gamma =3.3)/Z_{p{\rm -air}}^{c}(E,\gamma =3)$,
for the considered PDF sets. It is easy to see that
 a change of the slope of the primary
spectrum gives rise to a practically
 energy-independent rescaling of $Z_{p{\rm -air}}^{c}$.
This is due to the fact that such a change has a negligible effect on the
range of relevant $x^-$ values in Eqs.\ (\ref{eq:c-gg}-\ref{eq:sig-c}), 
while causing an additional suppression of small $x^+$ values, 
as illustrated in
 Fig.\ \ref{fig:z_c-xpm-gamma} for the CT14nlo\_NF3 PDF set. 
 It is noteworthy that the
 obtained dependence of $Z_{p{\rm -air}}^{c}$ on the primary spectrum slope is
 substantially weaker, compared to the corresponding dependence for  
 prompt neutrino fluxes (see, e.g.\ \cite{bha15}): since an additional (and
 stronger)  effect comes  in the latter case from the $\gamma$-dependence of the
 fragmentation and decay moments,\footnote{See, e.g.\ Table 3 in 
 Ref.\ \cite{tig96}.}
  $Z_{c\,(\bar c)\rightarrow h_c}^{\rm fragm}$
and $Z^{\rm dec}_{h_c\rightarrow \nu_{\mu}}$  
 [c.f., Eqs.\ (\ref{eq:final-fact}-\ref{eq:z-factors})].
 Interestingly, there are only minor differences between the values
of the ratio 
 $Z_{p{\rm -air}}^{c}(E,\gamma =3.3)/Z_{p{\rm -air}}^{c}(E,\gamma =3)$,
 obtained for the  considered PDF sets; the differences regarding the
 high-$x$ behavior of the respective gluon PDFs do not make any important
 impact on the $\gamma$-dependence of the $Z$-factors for charm production.
 
\begin{figure*}[t]
\includegraphics[width=\textwidth]{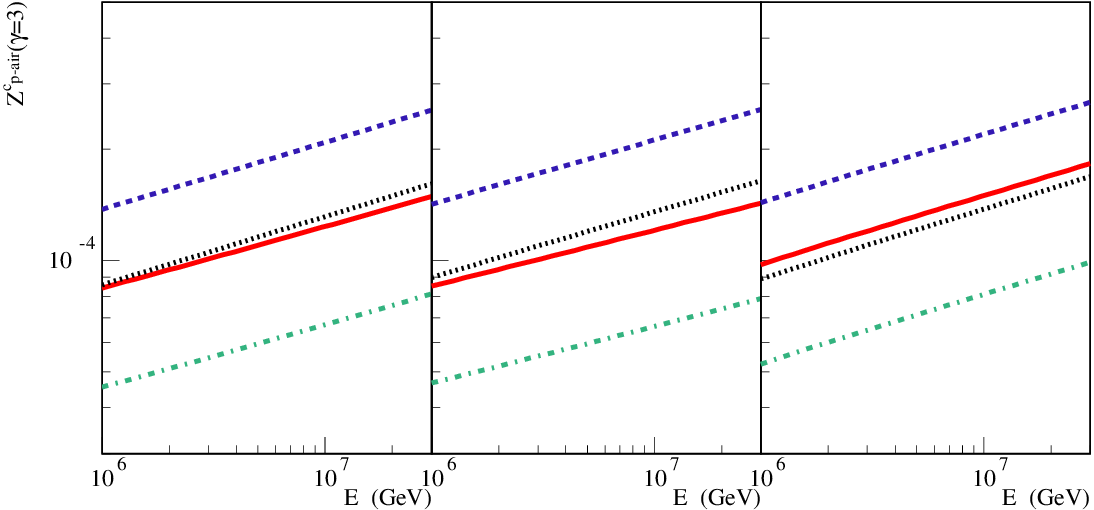}
\caption{CR spectrum-weighted moments of $c$-quark
production spectrum, $Z_{p{\rm -air}}^{c}(E,\gamma)$, for  $\gamma =3$,
calculated using different combinations of the factorization
  and  renormalization scales:  
  $(\mu_{\rm F},\mu_{\rm R})=(1,1)m_{\perp _c}$ (solid),
  $(\mu_{\rm F},\mu_{\rm R})=(2,1)m_{\perp _c}$  (dashed),
  $(\mu_{\rm F},\mu_{\rm R})=(1,2)m_{\perp _c}$  (dotted-dashed),
 and $(\mu_{\rm F},\mu_{\rm R})=(2,2)m_{\perp _c}$  (dotted).
 The graphs in the
 left, middle, and right panels are based on  gluon PDFs from
ABMP16\_3\_nlo, CT14nlo\_NF3, and NNPDF31\_nlo\_pch\_as\_0118\_nf\_3
  PDF sets, respectively.
\label{fig:scale-var}}
\end{figure*}

 Finally, in Fig.\ \ref{fig:scale-var},  we investigate the sensitivity of 
 calculated  CR spectrum-weighted moments $Z_{p{\rm -air}}^{c}$ to variations
 of the factorization
  $\mu_{\rm F}$ and  renormalization $\mu_{\rm R}$ scales: by comparing the
  respective results obtained with  $\mu_{\rm F}=\mu_{\rm R}=m_{\perp _c}$ to 
  the ones calculated using twice larger values  for  $\mu_{\rm F}$, or for
  $\mu_{\rm R}$, or for both. Clearly, the sensitivity to higher order
  pQCD corrections, reflected by the strong dependence of the results
 on the scale choices, represents the largest uncertainty regarding the 
 perturbative input for calculations of prompt neutrino fluxes, as already 
 stressed in previous studies \cite{zen20}. It is noteworthy that the uncertainty
 regarding the low-$x$ extrapolation of the gluon PDF had been greatly reduced
 by taking into consideration LHCb data on forward charm
  production \cite{gau15,gau17,zen15,gar17}.

\section{Intrinsic charm\label{IC.sec}}

Let us now discuss the non-perturbative contribution of the so-called
intrinsic charm \cite{bhps80,bps81}, which can potentially enhance prompt neutrino
fluxes \cite{lah17}. In some approaches, the corresponding charm production is
linked to interactions of constituent charm (anti)quarks from the respective
Fock states of the proton (e.g.\ $|uudc\bar c\rangle$) with a target gluon:
via the $cg\rightarrow cg$ hard scattering process (see, e.g.\ \cite{gia18,mac20}).
The picture one has in mind corresponds to a dense gluon cloud originating
from the target and incoming on the projectile proton, with some of these
gluons hitting the constituent charm (anti)quarks. In such a case, the 
corresponding contribution to  prompt neutrino fluxes is essentially
proportional to the gluon PDF of air nuclei, 
$g_{\rm air}(x^-,Q) \simeq \langle A_{\rm air}\rangle\,g_p(x^-,Q)$,
probed at very small values of the LC$^-$ momentum fraction 
$x^-\sim m_c^2/(x^+s)$
and relatively low $Q$. Consequently, one obtains the same 
$A$-enhancement of charm production  ($\propto \langle A_{\rm air}\rangle$), 
as for the perturbative
generation of charm [c.f., Eqs.\ (\ref{eq:c-gg}-\ref{eq:sig-c})] and, more importantly, the
same kind of   energy rise: 
$\propto\left. g_p(x^-,Q)\right|_{x^-\sim  m_c^2/(x^+s)}$.

What is missed in the above-discussed approaches is that at very high energies
we are interested in, the basic valence quark configuration is surrounded by
an extensive ``coat'' formed by gluons and sea quarks and, more importantly,
that this coat covers a substantially larger transverse area than the 
compact valence quark ``core'' \cite{fra04}. As a consequence, high energy
proton-proton (proton-nucleus) collisions are dominated by multiple scattering
processes between such non-valence partons and it is this multiple scattering
that unitarizes the respective interaction cross sections.

The above reasoning applies also to the case when the incoming proton is
represented by a constituent parton Fock state containing charm (anti)quarks,
like $|uudc\bar c\rangle$: at sufficiently high energies, these are 
gluons and sea quarks from the projectile proton, which typically
 interact with their
counterparts from the target. On the other hand, valence quarks usually stay
as ``spectators'' and participate in secondary particle production at the
hadronization stage only. Here the crucial point is that an interaction with
a non-valence constituent of the incoming proton is sufficient to destroy
the coherence of its original partonic fluctuation and thereby to ``free''
the charm quark-antiquark pair from its virtual state \cite{bro89,bro92}.  

Thus, at the energies of our interest, interactions of proton Fock states
containing intrinsic charm constitute a constant fraction $w^c_{\rm intr}$
of the inelastic cross section,\footnote{In contrast, in the low energy limit, 
the contribution of such states to the inelastic cross section is much
suppressed, compared to the basic   $|uud\rangle$ configuration. Indeed,
since their parton coat remains undeveloped, such states appear to be much
more compact than the   $|uud\rangle$ state \cite{bro89}.}
 with $w^c_{\rm intr}$ being the
overall weight of such states, as  suggested already in Ref.\ \cite{tig96}
(their model 1 for intrinsic charm). Consequently, the corresponding
contribution to charm (anti)quark production can be formally written as
\begin{equation}
\frac{d\sigma_{p{\rm -air}}^{c{\rm (intr)}}(E,x_c)}{dx_c}=w^c_{\rm intr}\,
\sigma_{p{\rm -air}}^{\rm inel}(E)\,f_c^{\rm (intr)}(x)\,,
\label{eq:sig-c-intr}
\end{equation}
with $f_c^{\rm (intr)}(x)$ being the (normalized to unity) distribution of the
constituent $c$-quark LC momentum fraction in the proton. The corresponding
CR spectrum-weighted moment is thus neither energy nor target mass 
dependent [c.f., Eq.\ (\ref{eq:X-gener})]:
\begin{eqnarray}
 Z_{p{\rm -air}}^{c{\rm (intr)}}(\gamma)= Z_{pp}^{c{\rm (intr)}}(\gamma)
&&\nonumber \\
=w^c_{\rm intr} \int \!dx_c\,x_c^{\gamma -1}\,f_c^{\rm (intr)}(x_c)\,.
\label{eq:z-intr}
\end{eqnarray}
For the particular case of $\gamma =3$, it is thus proportional to the second
moment of the constituent $c$-quark momentum distribution. An important 
consequence of the energy-independence of $Z_{p{\rm -air}}^{c{\rm (intr)}}$
is that the corresponding contribution to the prompt neutrino flux is
characterized by the same energy slope as the primary proton flux.

In Table~\ref{tab:z-intr}, we compare the calculated moments
$Z_{pp}^{c{\rm (intr)}}$ for two different distributions 
$f_c^{\rm (intr)}$ and for the primary proton spectral slopes $\gamma =3$
and $\gamma =3.3$. Our first choice corresponds to the original intrinsic
charm model of Brodsky-Hoyer-Peterson-Sakai (BHPS) \cite{bhps80,bps81}:
\begin{eqnarray}
f_c^{\rm (intr)}(x)\propto x^2\left[\frac{1}{3}(1-x)(1+10x+x^2)\right.
&&\nonumber \\
+\left. 2x(1+x)\ln x\right].
\label{eq:bps}
\end{eqnarray}

\begin{table}[t]
\begin{tabular*}{0.49\textwidth}{@{\extracolsep{\fill}}ccc}
\hline 
$\gamma$ & 3 & 3.3\tabularnewline
\hline 
\hline 
BHPS model & 0.0018 & 0.0014\tabularnewline
Regge model & 0.0020 & 0.0016\tabularnewline
\hline 
\end{tabular*}\caption{Calculated CR spectrum-weighted moments
$Z_{pp}^{c{\rm (intr)}}$, for the BHPS and Regge models of intrinsic charm,
using different primary spectral slopes.
\label{tab:z-intr}}

\end{table}

Alternatively, we consider a Regge ansatz:
\begin{equation}
f_c^{\rm (intr)}(x)\propto
x^{-\alpha_{\psi}}\,(1-x)^{-\alpha_{\psi}+2(1-\alpha_N)}\,.
\label{eq:regge}
\end{equation}
Here the factor $x^{-\alpha_{\psi}}$ corresponds to the probability to slow
down the constituent $c$-quark, with $\alpha_{\psi}\simeq -2$ being the
intercept of the $c\bar c$ Regge trajectory \cite{kai03}. On the other hand,
the limit $x\rightarrow 1$ is defined by the probability to slow down the
remaining (``dressed'') valence quark configuration ($uud\bar c$), which
contributes the factor $(1-x)^{-\alpha_{\psi}+2(1-\alpha_N)}$, with
$\alpha_N\simeq -0.5$ \cite{kai86}.

In both cases, we choose $w^c_{\rm intr}$ such that the total LC momentum
fraction of the proton, carried by $c$ and $\bar c$, equals 1\%, as suggested
by the global analyses of the proton PDFs by the CTEQ collaboration \cite{dul14}:
\begin{equation}
w^c_{\rm intr}=0.01/\langle x^{c+\bar c}_{uudc\bar c}\rangle = 0.01/[2\! \int\!
dx\;xf_c^{\rm (intr)}(x)].
\end{equation}

\begin{figure}[htb]
\includegraphics[width=0.49\textwidth]{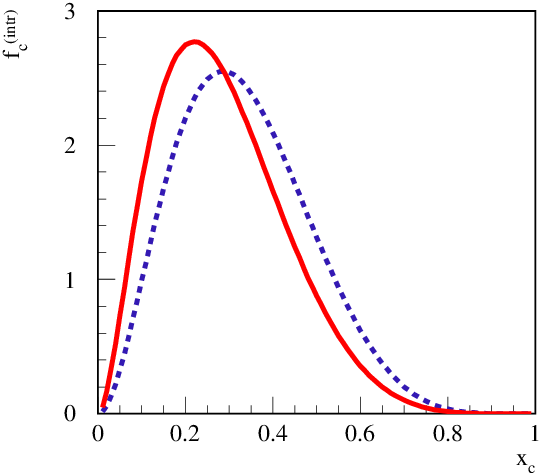}
\caption{Distribution of the
constituent $c$-quark LC momentum fraction in the proton, for the BHPS model
(solid line) and for the Regge ansatz (dashed line). \label{fig:f_c}}
\end{figure}

As we can see in Table~\ref{tab:z-intr}, the calculated moments 
$Z_{pp}^{c{\rm (intr)}}$ depend  weaker on the primary slope
$\gamma$ than  $Z_{p{\rm -air}}^{c}$ for perturbative charm production
(c.f., Fig.\ \ref{fig:z_c-slope}).
 This is not surprising  since for both our choices 
of $f_c^{\rm (intr)}(x)$ these distributions shown in Fig.\ \ref{fig:f_c}
 peak at larger values of $x$,
compared to  $dZ_{p{\rm -air}}^{c}/dx^+$ 
shown in Fig.\ \ref{fig:dz_c/dx}~(left).
Further, despite the fact that the fraction of proton LC momentum, carried by
charm (anti)quarks, is the same for our both models of intrinsic
charm, we obtained   somewhat larger  $Z_{pp}^{c{\rm (intr)}}$ 
when using the Regge ansatz, Eq.\ (\ref{eq:regge}). 
 This is because  that distribution is shifted towards
higher $x$ values, compared to the one of the BHPS model, while 
 the CR spectrum-weighted moments are  
proportional to the second moment of $f_c^{\rm (intr)}$ for $\gamma =3$
or to an even higher one for a steeper  CR spectrum.

If we formally compared the magnitudes of the  $Z$-factors from  
Table~\ref{tab:z-intr} to the ones
corresponding to perturbative charm production, plotted in Fig.\ \ref{fig:z_c},
we would come to the conclusion that even a sub-percent contribution
of intrinsic charm would be sufficient to dominate
the prompt  atmospheric flux of neutrinos.
However, such a comparison may be misleading  since the hadronization of 
constituent  charm (anti)quarks can proceed differently,
 compared to the perturbatively  generated ones, hence,  
 our reasoning in Section~\ref{form.sec}  may be inapplicable to the
  case of intrinsic charm. 
 Indeed, a  constituent $c$-quark is likely to recombine with
  a valence diquark of the proton to form a charmed baryon (e.g.\ 
  $c+ud\rightarrow \Lambda_c^+$), as suggested by  
  measurements of the $\Lambda_c$ production asymmetry
  by the SELEX experiment \cite{gar02}. In particular,
  such a picture is implicit in the so-called meson-baryon models of intrinsic
  charm \cite{mel14}. Therefore, a quantitative comparison of the
  perturbative and non-perturbative contributions to prompt neutrino fluxes
  can only be performed at the level of neutrino production, taking into
 consideration the differences between the hadronization mechanisms for
 the two cases, as was done
  in previous studies (see, e.g.\ \cite{lah17}). Nevertheless, the relatively
  large values of the   $Z$-factors listed in Table~\ref{tab:z-intr}  
  indicate that uncertainties regarding the potential non-perturbative
  contribution to prompt neutrino fluxes may dominate   the ones
  corresponding to perturbative charm production.

\section{Conclusions\label{summary.sec}}

In this work, we addressed the prompt contribution to the atmospheric neutrino
 flux. Concentrating on the particular case of muonic (anti)neutrinos, 
 we demonstrated
 that in the energy range of practical interest, the problem can be studied
  at the level of charm (anti)quark production. Indeed,  using the collinear 
  factorization framework of pQCD, we were able to conveniently factorize out
  both the  fragmentation functions for charm (anti)quarks and
the decay distributions for charmed hadrons, thereby expressing the 
 prompt flux of atmospheric neutrinos via CR spectrum-weighted
 moments ($Z$-factors)  of production spectra for   charm (anti)quarks.
 
 We illustrated the advantages of the method by studying the dependence of
 our results on the choice of gluon PDFs employed, on the value of the 
 primary CR spectral slope, and on the variations of the factorization and
 renormalization scales involved in the perturbative evaluation of charm
 (anti)quark production. We investigated also the range of 
 momentum fractions  of both projectile and target gluons,
which correspond to maximal contributions to  prompt atmospheric 
neutrino fluxes.

 Additionally, we discussed
the non-perturbative contribution to the prompt neutrino flux, related to
the intrinsic charm content of the proton, using two parametrizations  
for momentum distributions of constituent charm (anti)quarks in the proton.
We demonstrated
that the corresponding $Z$-factors take a particularly simple form, being
neither energy nor target mass number dependent in the energy range of interest.
Consequently,  the corresponding contribution to the prompt neutrino flux 
should be
characterized by the same energy slope as the primary CR flux.

However, our approach may be inapplicable for a quantitative comparison of the
  perturbative and non-perturbative contributions to prompt neutrino fluxes:
 because of potentially different hadronization mechanisms in   the two cases.
Nevertheless, it is worth stressing  that our observation regarding the energy-dependence of the contribution
of the intrinsic charm to the atmospheric neutrino spectrum, namely, that
it is characterized by the same spectral slope as the primary CR spectrum,
remains valid, regardless the hadronization mechanism. Since the corresponding
perturbative contribution is characterized by a flatter spectral slope
[c.f., Fig.\  \ref{fig:z_c} and Eq.\ (\ref{eq:flux-z})], 
this offers one a possibility to disentangle the
two contributions, based on IceCube data, once a sufficient experimental
statistics becomes available at the highest neutrino energies. 
 
\subsection*{Acknowledgments}

S.O.\ acknowledges enlightening discussions with S.\ Brodsky
and a financial support from  Deutsche Forschungsgemeinschaft 
(project number 465275045). M.V.G.\ is grateful to M.\ Benzke for 
useful cross-checks and to M.\ H.\ Reno
for discussions and clarifications on her works on related topics.
 The work of M.V.G.\ was partially supported by the Bundesministerium
  f\"ur Bildung und Forschung, under contract 05H21GUCCA. G.S.\ acknowledges
support by the Bundesministerium f\"ur Bildung
und Forschung, under grants 05A17GU1 and
05A20GU2.


\begin{thebibliography}{99}
\bibitem{aar13a} IceCube Collaboration, M.\ G.\ Aartsen et al., {\em First
observation of PeV neutrinos with IceCube}, Phys.\ Rev.\ Lett.\ {\bf 111} (2013)
021103.

\bibitem{aar13b} IceCube Collaboration, M.\ G.\ Aartsen et al., {\em Evidence
for High-Energy Extraterrestrial Neutrinos at the IceCube Detector},
 Science {\bf 342} (2013) 1242856.

\bibitem{gaisser-book}
T.\ K.\ Gaisser, {\em Cosmic Rays and Particle Physics} (Cambridge University
Press, Cambridge, England 1990).

\bibitem{lip93}
P.\ Lipari, {\em Lepton spectra in the earth's atmosphere},
Astropart.\ Phys.\ {\bf 1} (1993) 195.


\bibitem{bar04} G.\ D.\ Barr, T.\ K.\ Gaisser, P.\ Lipari, S.\ Robbins, and
T.\ Stanev, {\em A three-dimensional calculation of atmospheric neutrinos},
 Phys.\  Rev.\ D {\bf 70} (2004) 023006.

\bibitem{hon07} M.\ Honda, T.\ Kajita, K.\ Kasahara, S.\ Midorikawa, and T.\
Sanuki,  {\em Calculation of atmospheric neutrino flux using the interaction
model calibrated with atmospheric muon data},
 Phys.\  Rev.\ D {\bf 75} (2007) 043006.

\bibitem{tig96} M.\ Thunman, G.\ Ingelman, and P.\ Gondolo, {\em Charm
production and high energy atmospheric muon and neutrino fluxes},
Astropart.\ Phys.\   {\bf 5} (1996) 309.
 
 \bibitem{pas99} L.\ Pasquali, M.\ H.\ Reno, and I.\ Sarcevic,
 {\em Lepron fluxes from atmospheric charm}, 
 Phys.\  Rev.\ D {\bf 59} (1999) 034020.
 
 \bibitem{enb08} R.\ Enberg, M.\ H.\ Reno, and I.\ Sarcevic,
 {\em Prompt neutrino fluxes from atmospheric charm}, 
 Phys.\  Rev.\ D {\bf 78} (2008) 043005.
 
 \bibitem{bha15} A.\ Bhattacharya, R.\ Enberg, M.\ H.\ Reno,  I.\ Sarcevic,
 and A.\ Stasto,  {\em Perturbative charm
production and the prompt atmospheric neutrino flux in light of RHIC
and LHC},
 JHEP {\bf 06} (2015) 110.
 
 \bibitem{gar15} M.\ V.\ Garzelli, S.\ Moch, and G.\ Sigl,  {\em Lepton fluxes
from atmospheric charm revisited},
 JHEP {\bf 10} (2015) 115.

 \bibitem{gau15} R.\ Gauld, J.\ Rojo, L.\ Rottoli, and J.\ Talbert,
   {\em Charm production in the forward region: constraints on the small-$x$
   gluon and backgrounds for neutrino astronomy},
 JHEP {\bf 11} (2015) 009.
 
 \bibitem{bha16} A.\ Bhattacharya, R.\ Enberg, M.\ H.\ Reno,  I.\ Sarcevic,
 and A.\ Stasto,  {\em Prompt atmospheric neutrino fluxes: perturbative 
QCD models and nuclear effects},
 JHEP {\bf 11} (2016) 167.

 \bibitem{ben17} M.\ Benzke, M.\ V.\   Garzelli, B.\ Kniehl, G.\ Kramer,  
S.\ Moch,  and G.\ Sigl, 
 {\em Prompt neutrinos from atmospheric charm in the general-mass 
variable-flavor-number scheme},
 JHEP {\bf 12} (2017) 021.

\bibitem{kul59} G.\ Kulikov and G.\ Khristiansen, {\em On the size spectrum
of extensive air showers},
Sov.\ Phys.\ JETP  {\bf 8} (1959) 441.

\bibitem{ape13a} KASCADE-Grande Collaboration, W.\ D.\ Apel et al., 
{\em Ankle-like Feature in the Energy Spectrum of Light Elements of 
Cosmic Rays Observed with KASCADE-Grande},
 Phys.\  Rev.\ D {\bf 87} (2013) 081101.

\bibitem{ant05} KASCADE Collaboration, T.\ Antoni et al., {\em KASCADE
measurements of energy spectra for elemental groups of cosmic rays: 
Results and open problems},
Astropart.\ Phys.\   {\bf 24} (2005) 1.

\bibitem{ape13b} KASCADE-Grande Collaboration, W.\ D.\ Apel et al., 
{\em KASCADE-Grande
measurements of energy spectra for elemental groups of cosmic rays},
 Astropart.\ Phys.\   {\bf 47} (2013) 54.

\bibitem{ape11} KASCADE-Grande Collaboration, W.\ D.\ Apel et al., 
{\em Knee-like structure in the spectrum of the heavy component of 
cosmic rays observed with KASCADE-Grande},
 Phys.\  Rev.\ Lett.\ {\bf 107} (2011) 171104.

\bibitem{aar19} IceCube Collaboration, M.\ G.\ Aartsen et al., {\em Cosmic ray
spectrum and composition from PeV to EeV using 3 years of data from
IceTop and  IceCube}, Phys.\ Rev.\  D {\bf 100} (2019) 082002.

  
\bibitem{ko14} M.\ Kachelrie\ss~and S.\ Ostapchenko, {\em Neutrino yield
from Galactic cosmic rays},
 Phys.\  Rev.\ D {\bf 90} (2014) 083002.
 
\bibitem{kam12}   K.-H.\ Kampert and M.\ Unger, 
 {\em Measurements of the Cosmic Ray Composition with
Air Shower Experiments}, Astropart.\ Phys.\
 {\bf 35} (2012) 660.

\bibitem{pao21}    Pierre Auger Collaboration, P.\ Abreu et
al.,  {\em The energy spectrum of cosmic rays 
beyond the turn-down around $10^{17}$ eV as 
measured with the surface detector of the Pierre
Auger Observatory}, Eur.\ Phys.\ J.\ C {\bf 81}
(2021) 966.

\bibitem{eng92} J.\ Engel, T.\ K.\ Gaisser,  T.\ Stanev, and P.\ Lipari,
 {\em Nucleus-nucleus collisions and interpretation of cosmic ray cascades},
 Phys.\  Rev.\ D {\bf 46} (1992) 5013.
 
\bibitem{grv94} M.\ Gl\"uck, E.\ Reya, and M.\ Stratmann, {\em Heavy quarks
at high energy colliders},
 Nucl.\ Phys.\  {\bf B422} (1994)  37.

\bibitem{ost11} S. Ostapchenko,  {\em Monte Carlo treatment of hadronic 
interactions in enhanced Pomeron scheme: QGSJET-II model}, 
 Phys.\  Rev.\ D {\bf 83} (2011)  014018.
 
\bibitem{nas89} P. Nason, S.\ Dawson, and R.\ K.\ Ellis,
  {\em The one particle inclusive differential cross section for heavy
  quark production in hadronic collisions},
Nucl.\ Phys.\  {\bf B327} (1989)  49.
 
\bibitem{ct14} S.\ Dulat, T.-J.\ Hou, J.\ Gao, M.\ Guzzi,  J.\ Huston,  
P.\  Nadolsky, J.\ Pumplin,  C.\ Schmidt, D.\ Stump, and C.-P.\ Yuan,
  {\em New parton distribution functions from a global analysis of quantum 
chromodynamics},
 Phys.\  Rev.\ D {\bf 93} (2016) 033006.

\bibitem{ale12} S.\ Alekhin, J.\ Bl\"umlein,  and S.\ Moch,
  {\em NLO PDFs from the ABMP16 fit},
  Eur.\ Phys.\  J.\ C {\bf 78} (2018) 477.


\bibitem{bal11} NNPDF Collaboration,
 R.\ D.\ Ball, V.\ Bertone, F.\ Cerutti, S.\ Carrazza, L.\ Del Debbio,                                     
  S.\ Forte, P.\ Groth-Merrild, A.\ Guffanti, N.\ P.\ Hartland,
  Z.\ Kassabov et al.,
  {\em Parton distributions from high-precision collider data},
  Eur.\ Phys.\  J.\ C {\bf 77} (2017) 663.



\bibitem{lhapdf}  A.\ Buckley, J.\ Ferrando, S.\ Lloyd, K.\ Nordstr\"om,
 B.\ Page, M.\ R\"ufenacht, M.\ Sch\"onherr, and G.\ Watt,
 {\em LHAPDF6: parton density access in the LHC precision era},
 Eur.\ Phys.\  J.\ C {\bf 75} (2015) 132.
 
 \bibitem{zen20} PROSA Collaboration, O.\ Zenaiev,  M.\ V.\ Garzelli, K.\ Lipka, S.-O.\ Moch, 
A.\ Cooper-Sarkar, F.\ Olness, A.\ Geiser,  and G.\ Sigl,  {\em  Improved constraints
on parton distributions using LHCb, ALICE and HERA heavy-flavour measurements
and implications for the predictions for prompt atmospheric-neutrino fluxes},
 JHEP {\bf 04} (2020) 118.

 \bibitem{gau17} R.\ Gauld and J.\ Rojo, 
   {\em Precision Determination of the Small-$x$ Gluon from Charm Production at LHCb},
 Phys.\  Rev.\ Lett.\ {\bf 118} (2017) 072001.

 \bibitem{zen15} O.\  Zenaiev et al., PROSA Collaboration,
{\em Impact of heavy-flavour production cross sections measured 
by the LHCb experiment on parton distribution functions at low x},
 Eur.\ Phys.\  J.\ C {\bf 75} (2015) 396.

 \bibitem{gar17} M.\ V.\ Garzelli, S.\ Moch,  O.\  Zenaiev,
A.\ Cooper-Sarkar, A.\ Geiser, K.\ Lipka, R.\ Placakyte,  and G.\ Sigl, 
 PROSA Collaboration,
 {\em Prompt neutrino fluxes in the atmosphere with PROSA parton 
distribution functions},
 JHEP {\bf 05} (2017) 004.

\bibitem{bhps80} S.\ J.\ Brodsky, P.\ Hoyer, C.\ Peterson, and N.\ Sakai,
  {\em The Intrinsic Charm of the Proton},
 Phys.\  Lett.\ {\bf B93} (1980) 451.
 
\bibitem{bps81} S.\ J.\ Brodsky,  C.\ Peterson, and N.\ Sakai, {\em Intrinsic
heavy-quark states},
 Phys.\  Rev.\ D {\bf 23} (1981) 2745.

\bibitem{lah17} R.\ Laha and S.\ J.\ Brodsky, 
  {\em IceCube can constrain the intrinsic charm of the proton},
 Phys.\  Rev.\ D {\bf 96} (2017) 123002.

\bibitem{gia18} A.\ V.\ Giannini, V.\ P.\ Goncalves,  and F.\ S.\ Navarra,
 {\em Intrinsic charm contribution to the prompt 
atmospheric  neutrino flux},
 Phys.\  Rev.\ D {\bf 98} (2018) 014012.

\bibitem{mac20} R.\ Maciu\l{}a and A.\ Szczurek, 
  {\em  Intrinsic charm in the nucleon and charm production at large
  rapidities in collinear, hybrid and $k_T$-factorization approaches},
 JHEP {\bf 10} (2020) 135.

\bibitem{fra04} L.\ Frankfurt, M.\ Strikman, and C.\ Weiss,
  {\em Dijet production as a centrality trigger for $pp$ collisions at CERN
  LHC},
 Phys.\  Rev.\ D {\bf 69} (2004) 114010.
 
\bibitem{bro89} S.\ J.\ Brodsky and  P.\ Hoyer,
  {\em Nucleus as a Color Filter in QCD: Hadron Production in Nuclei},
 Phys.\  Rev.\ Lett.\ {\bf 63} (1989) 1566.
 
\bibitem{bro92} S.\ J.\ Brodsky,  P.\ Hoyer, A.\ H.\ Mueller, and W.-K.\ Tang,
  {\em New QCD production mechanisms for hard processes at large $x$},
Nucl.\ Phys.\  {\bf B369} (1992)  519.

\bibitem{kai03} A.\ B.\ Kaidalov, {\em $J/\psi$ $c\bar c$ Production in $e^+e^-$
and Hadronic Interactions},
JETP Lett.\ {\bf 77} (2003) 349.

\bibitem{kai86} A.\ B.\ Kaidalov and O.\ I.\ Piskunova, 
{\em Inclusive Spectra of Baryons in the Quark-Gluon String Model},
Z.\ Phys.\  C {\bf 30} (1986) 145.

\bibitem{dul14} S.\ Dulat, T.-J.\ Hou, J.\ Gao, J.\ Huston,  J.\ Pumplin,
C.\ Schmidt, D.\ Stump, and C.-P.\ Yuan,
  {\em  Intrinsic charm parton distribution functions from CTEQ-TEA global
  analysis},
 Phys.\  Rev.\ D {\bf 89} (2014) 073004.
 
\bibitem{gar02} SELEX Collaboration, F.\ G.\ Garcia et al.,
{\em Hadronic production of $\Lambda _c$ from 600 GeV/c $\pi ^-$, $\Sigma ^-$
and $p$ beams},
 Phys.\  Lett.\ {\bf B528} (2002) 49.
 
\bibitem{mel14} T.\ J.\ Hobbs, J.\ T.\ Londergan, and W.\ Melnitchouk,
{\em Phenomenology of nonperturbative charm in the nucleon},
 Phys.\  Rev.\ D {\bf 89} (2014) 074008.

 

\end{thebibliography}
\end{document}